\documentclass[usenatbib]{mn2e}  
\usepackage{graphicx}
\usepackage{graphics}
\usepackage{amssymb}
\usepackage{amsmath}
\usepackage{array}
\usepackage{bm}
\usepackage{natbib}
\usepackage[normalem]{ulem}

\def\refe@jnl#1{{#1}}
\def\aj{\refe@jnl{Astron.~J.}}                  
\def\araa{\refe@jnl{Annu.~Rev.~Astron.~Astrophys.}}
\def\apj{\refe@jnl{Astrophys.~J.}}                 
\def\apjl{\refe@jnl{Astrophys.~J.~Lett.}} 
\def\apjs{\refe@jnl{Astrophys.~J.~Suppl.}}         
\def\aap{\refe@jnl{Astron.~Astrophys.}}            
\def\mnras{\refe@jnl{Mon.~Not.~R.~Astron.~Soc.}}   
\def\prd{\refe@jnl{Phys.~Rev.~D}}        
\def\fcp{\refe@jnl{Fund.~Cos.~Phys.}}  
\def\physrep{\refe@jnl{Phys.~Rep.}}   
\def\physlett{\refe@jnl{Phys.~Lett.}}
\def\nat{\refe@jnl{Nat.}}

\def\ga{\mathrel{\mathpalette\fun >}}
\def\fun#1#2{\lower3.6pt\vbox{\baselineskip0pt\lineskip.9pt
\ialign{$\mathsurround=0pt#1\hfill##\hfil$\crcr#2\crcr\sim\crcr}}}

\def\rn{} 
\def\nnn#1#2 #3{#2. #3. #1}                    
\def\nnnn#1 #2 #3 #4{#2. #3. #4. #1}            
\def\nnnnn#1 #2 #3 #4 #5{#2. #3. #4 #5. #1}     
  
\def\rf#1;#2;#3;#4;#5 {{\frenchspacing\par\rn#1, #3 {\bf #4}, #5 (#2). \par}} 
\def\rfbook#1;#2;#3;#4;#5 {{\frenchspacing\par\rn#1, {\it #3} (#5, #4,#2).\par}} 
\def\rfprep#1;#2;#3 {{\par\frenchspacing\rn#1, #3 (#2).\par}}

\newcommand{\nc}{\newcommand}

\nc{\beq}[1]{\begin{equation}\label{#1}} \nc{\eeq}{\end{equation}}

\nc{\inv}[1]{\frac{1}{#1}}

\def\gsim{\; \raise0.3ex\hbox{$>$\kern-0.75em
\raise-1.1ex\hbox{$\sim$}}\; }

\nc{\hn}{\hat{\textbf{n}}}
\nc{\bfi}[1]{\textit{#1}}

\begin{document}


\title[Cross-correlation between the cosmic microwave and infrared backgrounds for ISW detection]{Cross-correlation between the cosmic microwave and infrared backgrounds for integrated Sachs-Wolfe detection}


\author[Ili\'c, Douspis, Langer, P\'enin \& Lagache]{St\'ephane Ili\'c, Marian Douspis, Mathieu Langer, Aur\'elie P\'enin, Guilaine Lagache \\
Institut d'Astrophysique Spatiale, UMR8617, Universit\'e Paris-Sud \& CNRS, B\^at. 121, Orsay F-91405, France}

\date{}

\maketitle  

\begin{abstract} 
We investigate the cross-correlation between the cosmic infrared background (CIB) and cosmic microwave background (CMB) anisotropies due to the integrated Sachs-Wolfe (ISW) effect. We first describe the CIB anisotropies using a linearly biased power spectrum, valid on the angular scales of interest. From this, we derive the theoretical angular power spectrum of the CMB-CIB cross-correlation for different instruments and frequencies. Our cross-spectra show similarities in shape with usual CMB/galaxies cross-correlations. We discuss the detectability of the ISW signal by performing a signal-to-noise (SNR) analysis with our predicted spectra. Our results show that~: (\bfi{i}) in the ideal case of noiseless, full-sky maps, the significances obtained range from $6$ to $7\sigma$ depending on the frequency, with a maximum at 353 GHz (\bfi{ii}) in realistic cases which account for the presence of noise including astrophysical contaminents, the results depend strongly on the major contribution to the noise term. They span from $2$ to $5\sigma$, the most favorable frequency for detection being 545~GHz. We also find that the joint use of all available frequencies in the cross-correlation does not improve significantly the total SNR, due to the high level of correlation of the CIB maps at different frequencies.
\end{abstract} 



\maketitle


\section{Introduction}
\label{sec:introduction}

The discovery of the acceleration of the expansion of the Universe, made through supernov{\ae} observations \citep{SN1,SN2} at the end of the last century, has since led to many theories aimed at explaining its origin. These theories have been regrouped under the term ``Dark Energy" (DE), designating a new and unknown component of our Universe which theoretically amounts to 70\% of its total energy budget. Among the many solutions proposed to account for this intriguing phenomenon, one of the leading contenders is the so-called ``cosmological constant", an idea first introduced by Einstein in his original theory of general relativity to achieve a stationary universe, but which he discarded after the discovery of the Hubble redshift. This cosmological constant is assimilated to an intrinsic energy density of the vacuum, and therefore is constant in time and space ; it also has an equation of state $w=p/\rho$ equal to $-1$, both on theoretical grounds and because no confirmed deviations from $w=-1$ have been detected so far. Despite its simplicity, it does reproduce most of the current observations while being (one of) the most ``economical" solution, but it is nevertheless plagued by a few serious theoretical problems \citep[e.g.][]{Padmanabhan2003, Bass2011a}. \\

\newpage
 
Apart from these theoretical issues, the accelerated expansion of the Universe still needs to be tested further in the framework of the $\Lambda$-CDM model by independent measurements from cosmological observations. Over the last decade, other possible probes have been proposed such as the study of baryon acoustic oscillations \citep[][and references therein]{BAO1, BAO2, BAO3} which provide a ``standard ruler" in cosmology, and are heavily influenced by the energy content of the Universe --~and so, by the dark energy. \\

In this article, we focus on an alternative probe of the Dark Energy, namely the study of the integrated Sachs-Wolfe (ISW) effect ; the ``original" SW effect first introduced at the end of the 60s \citep{SW67} describes the imprint on the cosmic microwave background (CMB) of anisotropies caused by gravitational redshift occurring at the surface of last scattering. Its ``integrated" counterpart is similar in that it also has a gravitational origin and contributes to the CMB secondary anisotropies, but it only occurs in a Universe not dominated by matter. Indeed, the ISW effect is caused by the large-scale structures of the Universe, whose gravitional potentials are slowly decaying --~instead of being constant in a matter-dominated regime~-- and therefore giving a net gain (in case of a potential well) or loss (hill) of energy to the CMB photons that travel across them. \\

\newpage

This effect shows in the power spectrum of the CMB temperature anisotropies at large angular scales (low $\ell$) but the cosmic variance at those very multipoles together with the relatively small amplitude of the ISW effect make its direct detection very challenging, if not impossible, when using only the CMB itself. To circumvent this limitation, cosmologists have devised a way to exploit the link between this imprint on the CMB and the large-scale structures causing it, by simply cross-correlating the CMB with matter density maps (actually galaxy maps in practice) and then comparing the results to a null hypothesis and to what is expected from theory. \\

During the last decade or so, a growing interest has arisen in this field thanks to the development of large galaxy surveys~: SDSS \citep[][for the latest release]{SDSS}, NVSS \citep{NVSS}, 2MASS \citep{2MASS}, etc. They allow cosmologists to cross-correlate the CMB (as seen by WMAP) to proxies of the matter density field as seen at many wavelengths : X-rays \citep{ISW_Xray}, optical \citep{ISW_opt}, near-infrared \citep{ISW_NIR} or radio \citep{ISW_radio}. However, this method has yet to produce a definitive and conclusive detection of the ISW effect, with significances so far ranging from negligible \citep{ISW_negl} to 4.5$\sigma$ \citep{ISW_Gian} throughout the literature. The potential of future surveys such as LSST, Pan-STARRS or Euclid, have also been explored in terms of signal-to-noise ratio of the ISW detection \citep{DCCA}. Another noteworthy approach by \citet{ISW_tSZ} considered the cross-correlation of the ISW effect with the thermal Sunyaev-Zel'dovich effect as both effects take place in the same potential wells ; this could provide an independent probe for the existence of Dark Energy out of pure CMB data.\\

The originality of our work is to consider here the Cosmic Infrared Background (CIB), first discovered by \citet{CIB_Puget}. This background, visible roughly from $10$ to $1000~\mu m$ in wavelength, arises from accumulated emissions from star-forming galaxies spanning a large range of redshifts. The earliest epoch for the production of this background is thought to be when star formation first began, and contributions to the CIB continued through the present epoch, including our current dark energy dominated era. The CIB also features anisotropies \citep[first detected and discussed in][]{first_anis,first_anis2} that are underlined by the galaxy density field and thus the matter density fluctuations. It is therefore reasonable to expect that it has a positive correlation with the CMB through the aforementioned ISW effect.\\

   \begin{figure*}
   \centering
   \includegraphics[width=16cm]{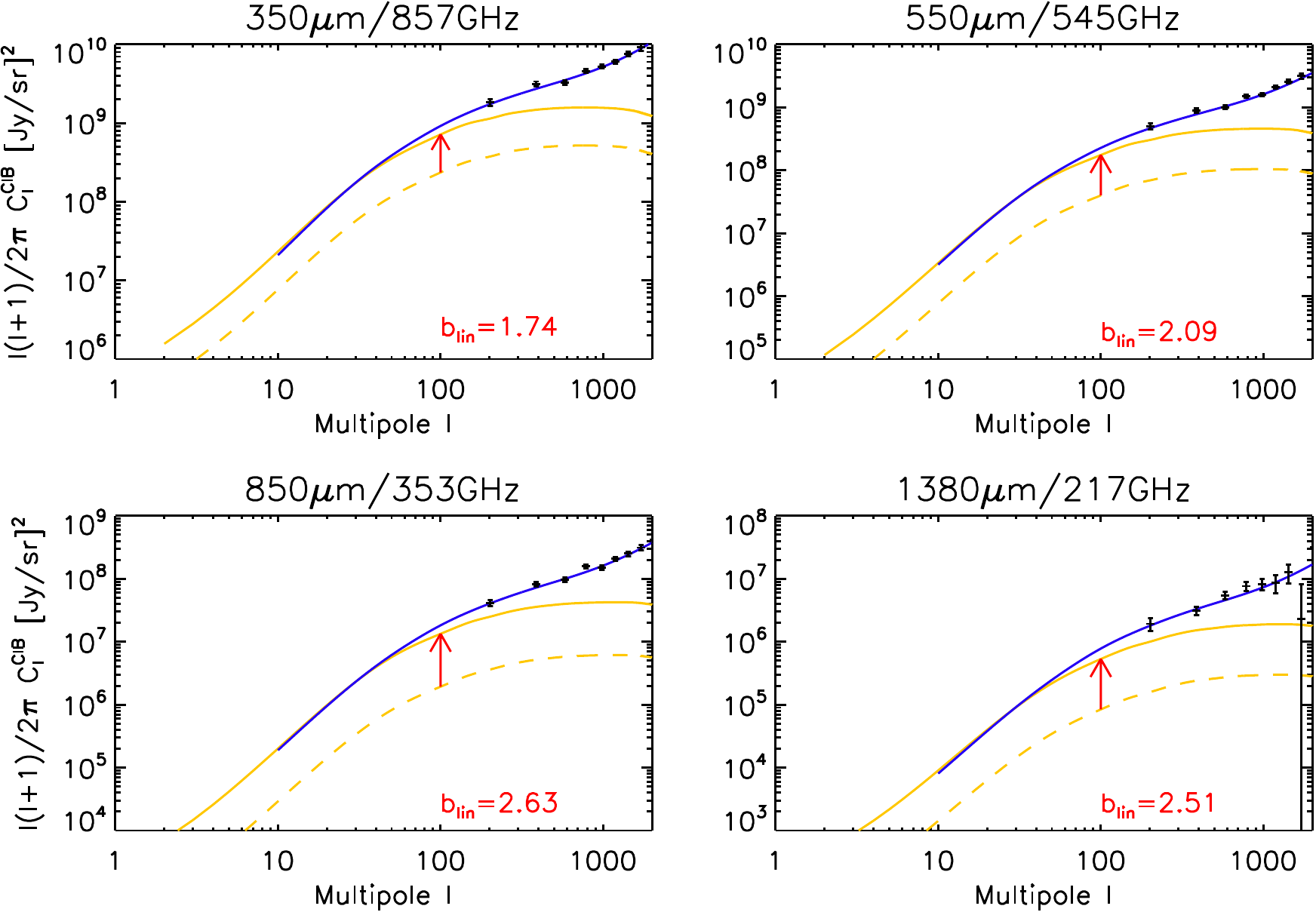}
 	 \caption{Angular power spectra of the CIB fluctuations at four frequencies of the \textit{Planck}-HFI instrument, as predicted by the \textit{Planck} Team (blue continuous line) and by our non-biased models (dashed yellow line). For each frequency, we provide in red the linear bias which gives the best agreement between the two models, and plot our models taking into account this bias (solid yellow line). The data points correspond to measurements obtained by the \textit{Planck} Team \citep{CIB_Planck}.}
         \label{compar_cl}
   \end{figure*}

In this paper, we first present an analytical calculation of the CMB-CIB cross-correlation signal through the integrated Sachs-Wolfe effect. We then use it to compute the expected power spectrum of this correlation in different cases, namely with a CIB observed at several frequencies and with various instruments (IRAS, \textit{Herschel}-SPIRE, and \textit{Planck}-HFI). With these results we perform a signal-to-noise ratio (SNR) analysis in order to quantify the detectability of the cross-correlation, focusing first on the ``perfect case" scenario, i.e. a situation where both the CMB and CIB are full sky maps, without noise, so that the detection is only limited by the cosmic variance. We then discuss the effect of noise (including contaminating astrophysical components and instrumental noise) in the maps and its consequences on the SNR. Finally, we end with a few conclusions and discussions about the perspective of application of our predictions.\\

Throughout all our calculations we assume a Euclidean Universe corresponding to the WMAP 7 best-fit cosmology, with adiabatic scalar perturbations and a nearly scale invariant initial power spectrum.


\section{Modelling the expected signal}
\label{sec:cib_isw}


\subsection{CIB anisotropies}
\label{sec:cib_anis}

Ever since its discovery, many efforts have been deployed to detect the cosmic infrared background with increasing precision, especially in order to study its anisotropies which contain a lot of information about the star and galaxy formation histories, including their clustering processes. The most recent papers on the CIB anisotropies use sophisticated models which compute the halo occupation distribution \citep[HOD, see e.g.][]{HOD1, HOD2} and the Dark Matter halos properties, in order to predict the power spectrum of these anisotropies. Recently applied to the new \textit{Planck} data \citep[see][]{CIB_Planck}, this framework allowed us to confirm that the bias between infra-red galaxies and the linear theory matter power spectrum is not independent of scale, and that the halo occupation distribution is evolving with redshift. \\

Such models are particularly useful when describing the small, non-linear scales of the CIB. Since we focus here on the ISW effect which only concerns much larger scales, we can use a simpler model for the CIB, similar to the description made by \citet{Knox01}. 
The general definition of the CIB anisotropies at a given frequency $\nu$ and in a given direction $\hn$ can be then written as the following line-of-sight integral~:
\beq{dtcib0}
\delta T_{\rm{CIB}}(\hn,\nu)=\int_{\eta_{\rm{far}}}^{\eta_0} dz \, \frac{d\eta}{dz} \, a(z) \, \delta j((\eta_0-\eta)\hn,\nu,z) \ 
\eeq
with $\delta j$ being the emissivity fluctuations of the CIB. The integration is made over $\eta$, the conformal time, from some initial time $\eta_{\rm{far}}$ before star formation began to our location at the coordinate origin $\eta_0$. In their work, Knox et al. hypothesized that the CIB anisotropies are direct tracers of the matter density fluctuations $\delta=\delta\rho_m/\bar{\rho}_m$, up to a bias factor. Therefore, the previous expression becomes an integral of the product between a mean far infrared (FIR) emissivity and the matter density fluctuation field~:
\beq{dtcib}
\delta T_{\rm{CIB}}(\hn,\nu)=\int_{\eta_{\rm{far}}}^{\eta_0} \! dz \, \frac{d\eta}{dz} \, a(z) \, b_j(\nu,z) \, \bar{j}(\nu,z) \, \delta((\eta_0-\eta)\hn,z) \ 
\eeq
\noindent The quantity $b_j(\nu,z)$ is a frequency- and redshift-dependent matter-emissivity bias defined by~:
\beq{biais}
\frac{\delta j((\eta_0-\eta)\hn,\nu,z)}{\bar{j}(\nu,z)} = b_j(\nu,z) \ \delta((\eta_0-\eta)\hn,z)
\eeq
and $\bar{j}(\nu,z)$ is the mean emissivity per comoving unit volume at frequency $\nu$ as a function of redshift $z$, which is derived here using the empirical, parametric model of \citet{Beth}. The matter density field $\delta$ is described in our analysis by a linear power spectrum. While this approximation is not accurate at small scales where non-linearities arise, it is perfectly valid for the scales of interest in our work since the first hundred multipoles ($\ell<100$) comprise most of the ISW signal. Following a calculation similar to the one described in the next section, we can express the angular power spectrum of the CIB fluctuations as follows~:
\beq{CIBpow}
C_{\ell}^{\rm{CIB}}(\nu) = 4\pi \frac{9}{25} \int \! \frac{dk}{k} \Delta_{\mathcal{R}}^2 \, M_{\ell}^2(k,\nu)
\eeq
where $M_{\ell}(k,\nu)$ is given in Eq.~(\ref{km}) and $\Delta_{\mathcal{R}}^2$ is defined below Eq.~(\ref{cross_cl}). \\ 

   \begin{figure*}
   \centering
   \includegraphics[width=14cm]{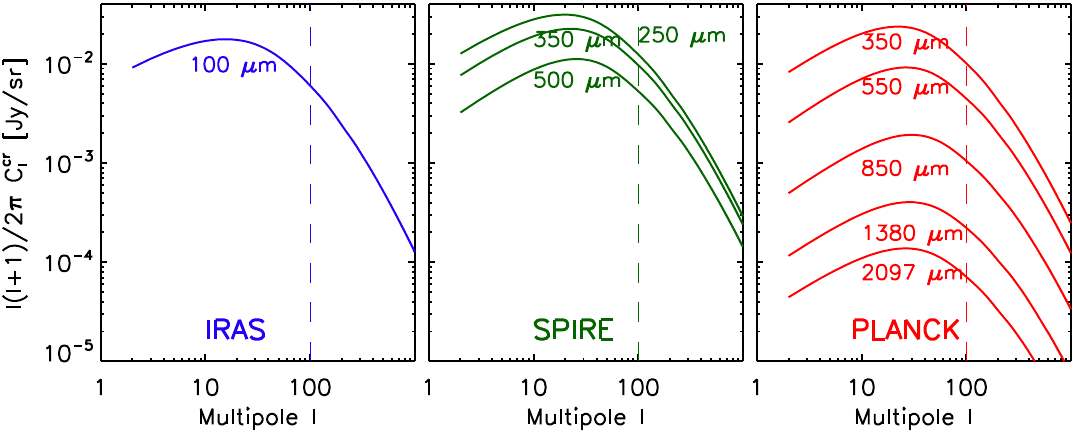}
      \caption{Theoretical angular cross power spectrum of the CIB-CMB correlation calculated for IRAS at 100~$\mu m$ (left-hand panel), for \textit{Herschel} SPIRE between 250-500~$\mu m$ (central panel) and \textit{Planck} HFI between 350-2097~$\mu m$ (right panel). The linear bias, $b_{\, \rm{lin}}$, is fixed here to 1 at all frequencies in order to compare the non-biased CIB power spectra. The vertical dashed line on each panel marks the upper limit of the multipoles used in our analysis~: this choice comes from the absence of ISW signal (see Fig.~\ref{plotsnr}) and the rise of non-linearities at higher $\ell$.}
         \label{cross}
   \end{figure*}

\newpage

Lastly we choose the previously mentioned linear bias~\footnote{This bias here represents our matter-emissivity bias in Eq.~(\ref{biais}) and should not be confused with the widely used galaxy-Dark Matter bias, though ours \textit{does} contain information about how the emitting objects populate Dark Matter halos.}
$b_j(\nu,z)$ to be constant in redshift~:  $b_j(\nu,z) \! = \! b_{\, \rm{lin}}(\nu)$. To obtain it at each frequency, we compute the value of $b_{\, \rm{lin}}$ that gives the best agreement between our linear CIB power spectrum and those obtained from the \textit{Planck} data \citep{CIB_Planck}. We choose to fit the two spectra in the range of multipoles $\ell \in [10,50]$, where most of the ISW signal resides. This is illustrated in Fig.~\ref{compar_cl} where we plot the biased and non-biased CIB linear spectra from our framework and compare them to the ones from \citet{CIB_Planck} at their four frequencies. Overall, the two sets of spectra show good agreement over the multipoles of interest ; the spectra deviate at higher $\ell$ (starting from $\simeq 100$) due to the rise of non-linearities that we did not account for in our linear model --~namely the small-scale correlations between galaxies inside the same halos. The linear bias we obtain this way increases with the wavelength~: this is coherent with the fact that as we go further deep into the infrared, the galaxies probed are more luminous at higher $z$. They reside in more massive and rarer halos, and are therefore more biased.


\subsection{Correlation with the ISW}

In the CMB anisotropies, the temperature contribution due to the ISW effect is an integral over the conformal time of the growth rate of the gravitational potentials~:

\beq{dtisw}
\delta T_{\rm{ISW}}(\hn)=\int_{\eta_r}^{\eta_0} d\eta \ e^{-\tau(\eta)} \ (\dot{\Phi}-\dot{\Psi})[(\eta_0-\eta)\hn,\eta]
\eeq
where $\eta_r$ is some initial time deep in the radiation era, $\Phi$ and $\Psi$ are the Newtonian gauge gravitational potentials \citep[with the conventions used in][]{gauge}, $\tau(\eta)$ is the optical depth included to account for the possibility of late reionisation, and the dot denotes differentiation with respect to $\eta$. \\

We are interested in calculating the CIB-ISW cross-correlation function $C^{\rm{cr}}$ at a given frequency $\nu$ in direct space~:
\beq{cross_real}
C^{\rm{cr}}(\theta_{\hn_1,\hn_2},\nu)  \equiv \langle \, \delta T_{\rm{CIB}}(\hn_1,\nu) \, \delta T_{\rm{ISW}}(\hn_2) \, \rangle.
\eeq
After a decomposition into Legendre series, we get :
\beq{cross_leg}
C^{\rm{cr}}(\theta,\nu) = \sum_{l=2}^{\infty} \frac{2\ell+1}{4\pi} \, C_{\ell}^{\rm{cr}}(\nu) \, P_{\ell}(\cos \theta)
\eeq
where we do not include the monopole and dipole terms in the sum. Using Eqs.~(\ref{dtcib}) and (\ref{dtisw}), we follow a calculation similar to \citet{Doom} in order to finally get the CMB-CIB cross-power spectrum at a frequency $\nu$~:
\beq{cross_cl}
C_{\ell}^{\rm{cr}}(\nu) = 4\pi \frac{9}{25} \int \! \frac{dk}{k} \Delta_{\mathcal{R}}^2 \, T_{\ell}^{\rm{ISW}}(k) \, M_{\ell}(k,\nu)
\eeq
where $\Delta_{\mathcal{R}}^2$ comes from the primordial curvature power spectrum $P_{\mathcal{R}} \equiv 2\pi^2\Delta_{\mathcal{R}}^2/k^3$. The use of this primordial spectrum differs from previous works on CMB-galaxies cross-correlation, where the present matter power spectrum is usually introduced instead, and is then evolved backward in order to find its correlation with the CMB. Conversely, in the Garriga et al. approach, the starting point is the primordial perturbations which are evolved to the present time. While it allows a full account of possible fluctuations in the dark energy in non-$\Lambda$ models, it also avoids the frequently used Limber approximation, which is known to be somewhat inaccurate at the largest scales considered here. \\

   \begin{figure*}
   \centering
   \includegraphics[width=14cm]{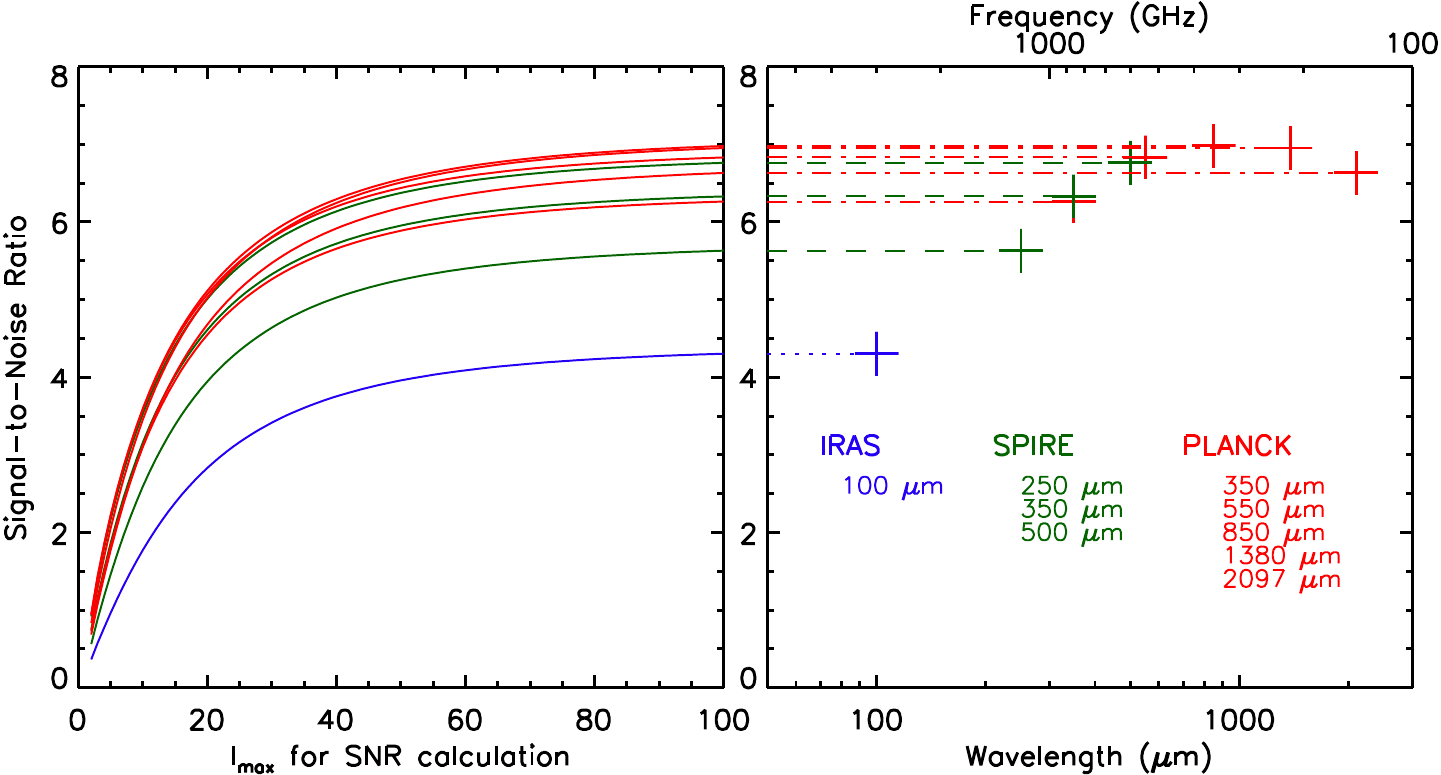}
      \caption{\textit{Left-hand panel :} Cumulated SNR as a function of $\ell_{max}$ (defined in Eq.~\ref{sn1}) for the CMB-CIB cross-correlation, at our chosen frequencies and instruments. \textit{Right-hand panel :} Total SNR with $\ell_{max}=100$ as a function of frequency/wavelength.}
         \label{plotsnr}
   \end{figure*}

At this point, we need to compute the two main functions $T_{\ell}^{\rm{ISW}}$ and $M_{\ell}$, which are defined as~:
\beq{kisw}
T_{\ell}^{\rm{ISW}}(k) = \int_{\eta_0}^{\eta_r} \! d\eta \ e^{-\tau(\eta)} \ j_{\ell}(k[\eta-\eta_0]) \ (c_{\Psi\Phi}\dot{\psi}-\dot{\phi}),
\eeq
and
\beq{km}
M_{\ell}(k,\nu) = c_{\delta\Psi} \! \! \int_{\eta_0}^{\eta_r} \! d\eta \ j_{\ell} \, (k[\eta-\eta_0]) \, a(\eta) \, b_{\rm{lin}}(\nu) \, \bar{j}(\nu,\eta) \, \tilde{\delta}(k,\eta)
\eeq
where $j_{\ell}(\cdot)$ are the spherical Bessel functions, while $\tilde{\delta}$, $\phi$ and $\psi$ are the time-dependent
\footnote{The separation between time and space dependance in the terms $\delta$, $\Phi$ and $\Psi$ is allowed in our calculation since the time evolution of each Fourier mode only depends on the magnitude~$k=||\textbf{k}||$. For exemple~: $\Phi(\textbf{k},\eta)=\Phi(\textbf{k},\eta_r)\phi(k,\eta)$ .} 
parts of (respectively) the Dark Matter density contrast $\delta$, and the two Newtonian gravitational potentials $\Phi$ and $\Psi$. The two coefficients $c_{\Psi\Phi}$ and $c_{\delta\Psi}$ give the relations between $\delta$, $\Phi$ and $\Psi$ for adiabatic initial conditions~:
\beq{coeffs}
c_{\delta\Psi} \equiv \frac{\delta}{\Psi}=-\frac{3}{2} \, , \ \ c_{\Psi\Phi} \equiv \frac{\Psi}{\Phi} = -\left( 1 + \frac{2}{5} R_\nu \right)
\eeq
where $R_\nu \equiv \rho_\nu/(\rho_\nu+\rho_\gamma)$, $\rho_\nu$ and $\rho_\gamma$ being respectively the energy densities in relativistic neutrinos and photons.


\subsection{Shape of the cross-correlation spectrum}
\label{sec:shape}

To compute these expressions, we adapted for our analysis an already modified version of CMBFAST \citep{CMBFAST}, named CROSS\_CMBFAST \citep{CCMBF}. For a given cosmology and emissivity function $\bar{j}(\nu,z)$ (see Eqs.~(\ref{dtcib}) and (\ref{biais})), our code calculates the $C_{\ell}^{\rm{cr}}$ from Eq.~(\ref{cross_cl}) and at the same time the predicted power spectrum of the CIB fluctuations described in Eq.~(\ref{CIBpow}) and already illustrated in Fig.~\ref{compar_cl}. It also gives the standard CMBFAST outputs, including the CMB temperature power spectrum.\\

In Fig.~\ref{cross}, we present our predictions for the CIB-CMB cross-correlation, at several FIR wavelengths and for different instruments, namely : IRAS at 100~$\mu m$, \textit{Herschel} SPIRE at 250, 350 and 500~$\mu m$ and \textit{Planck} HFI at 350, 550, 850, 1380, and 2097~$\mu m$. We note that at 350~$\mu m$ the SPIRE- and \textit{Planck}- predicted spectra differ slightly from each other, due to the difference in wavelength bandwidth of the two instruments. \\

In a fashion similar to previous galaxy-ISW cross-correlations (see the references in Section~\ref{sec:introduction}), we note that the cross-correlation peaks around $\ell\simeq10\textendash30$, and quickly vanishes at higher multipoles. Comparing the signal at the different wavelengths shows that the amplitude of the cross-correlation signal is maximum at a wavelength $\simeq 250~\mu m$. This is not entirely surprising, since this wavelength roughly corresponds to the maximum of the observed CIB spectral energy distribution \citep[see][for reference]{CIB_SED}. \\

It should be also noted that these results are not exact at the highest $\ell$s since the non-linear counterpart to the ISW effect, called the Rees-Sciama effect, contributes at those scales \citep[see][for a discussion]{RSeffect}. However, in our case the linear part of the ISW largely dominates at the observed peak in Fig.~\ref{cross}.


\section{Signal-to-Noise analysis}
\label{sec:sn}

\subsection{Ideal case}
\label{sec:perf_sn}

We now investigate the detection level of the ISW effect using CMB-CIB cross-correlation by performing a signal-to-noise ratio analysis. Using the power spectra computed in the previous section, we can write for each given frequency $\nu$ the total signal-to-noise ratio of the ISW detection as :
\beq{sn1}
\left[\frac{S}{N}\right]^2 \! \! \! \! (\nu) = \sum_{\ell=2}^{\ell_{\rm{max}}} (2\ell+1)\frac{[C_{\ell}^{\rm{cr}}(\nu)]^2}{[C_{\ell}^{\rm{cr}}(\nu)]^2+C_{\ell}^{\rm{CIB}}(\nu) \times C_{\ell}^{\rm{CMB}}}
\eeq
where the total (or cumulative) signal-to-noise is summed over multipoles between $\ell=2$ and $\ell_{\rm{max}} \leqslant 100$ where the signal has its major contribution (see previous section, Fig.~\ref{cross}). \\

In this section, we first consider the ideal situation where the CIB and CMB maps used for cross-correlation are noiseless and cover the whole sky ; with these assumptions we obtain the highest possible signal-to-noise ratio, the only limitation being the cosmic variance. In Fig.~\ref{plotsnr} we present our prediction for the CIB-CMB cross-correlation in the case of a full-sky CIB map, provided\footnote{Only the IRAS $100~\mu m$ data is already available, and previous works have managed to extract its CIB component on small patches of sky \citep*{CIB_IRAS}, but the CIB has yet to be extracted over a large enough part to allow for an ISW detection.}
by the previously mentionned instruments and frequencies. \\

With these optimistic assumptions, we obtain high levels of detection for the CIB-CMB correlation which reach $\simeq7\sigma$ (for detailed results, see Table~\ref{SNRtab}). It should be mentioned that these results in the ideal case are independant of the previously discussed linear bias in Section~\ref{sec:cib_anis} even if it boosts the correlation signal. This can be understood from Eq.~(\ref{sn1}) where the linear bias can be factorized from each term (one for $C_{\ell}^{\rm{cr}}$ and a squared one for $C_{\ell}^{\rm{CIB}}$) and therefore cancels out. \\

\begin{figure*}
   \centering
   \includegraphics[width=\textwidth]{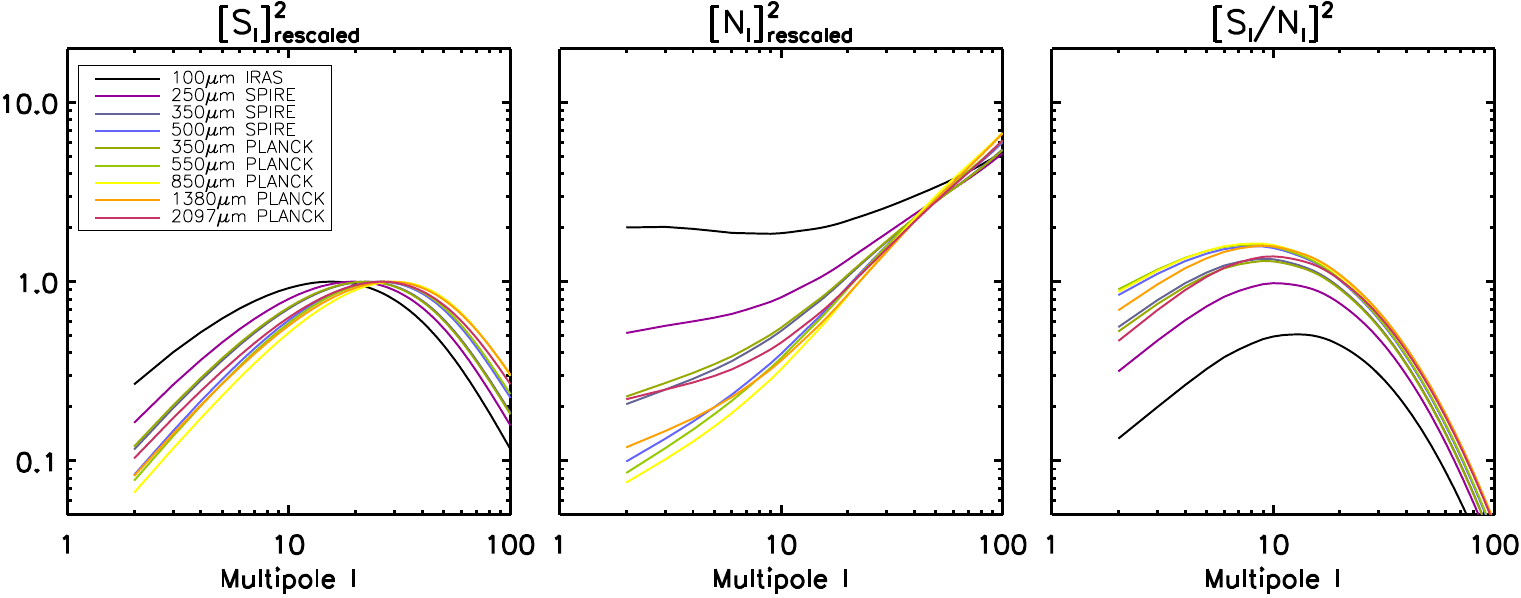}
      \caption{``Signal" terms (left-hand panel, rescaled to unity) and ``noise" terms (middle panel, same rescale factor as the ``signal") of the SNR as functions of $\ell$ (see text for details) for our chosen frequencies and instruments.  The quotient of the two terms, used in the calculation of the SNR itself, is shown in the right-hand panel~: the main difference throughout the frequencies comes from the shape of the ``noise" term.}
         \label{explsnr}
   \end{figure*}

As evoked in Section \ref{sec:shape}, we see that the largest contribution to the SNR comes from multipoles lower than $\simeq50$. On the other hand, the most interesting feature of these results is that contrary to what could be intuited from Fig.~\ref{cross}, the total SNR peaks around 850~$\mu m$ instead of 250~$\mu m$ for the cross-correlation signal itself. The reason for this is actually quite subtle~: it comes from the shape of the ``noise" term in the SNR expression in Eq.~(\ref{sn1}), as a function of $\ell$, namely~:
\begin{equation*}
[N_\ell]^2(\nu) \equiv ([C_{\ell}^{\rm{cr}}(\nu)]^2+C_{\ell}^{\rm{CIB}}(\nu) C_{\ell}^{\rm{CMB}})/(2\ell+1)
\end{equation*} 
For all the frequencies studied here, this ``noise" has roughly the same amplitude \textit{relatively} to its corresponding ``signal"~:
 \begin{equation*}
[S_\ell]^2(\nu) \equiv [C_{\ell}^{\rm{cr}}(\nu)]^2
\end{equation*} 
This is illustrated in Fig.~\ref{explsnr}, where we plotted in the left panel all the $[S_\nu]^2$ terms with their respective maximum rescaled to unity. In the middle panel, we apply the same rescaling factor of each $[S_\nu]^2$ term to the corresponding $[N_\nu]^2$ term. By doing this, we can compare all frequencies without changing their associated signal-to-noise ratios. On the resulting graph, we see that at $\ell=100$ the rescaled noise amplitude is roughly the same, while the signal has the same shape at all frequencies, except for a small shift in $\ell$. However there is a major difference in the shape of the noise power spectrum from one frequency to another~: its slope changes depending on the frequency, with the steepest one for \textit{Planck} 850~$\mu m$. Therefore its amplitude goes down more quickly than the others as $\ell$ approaches zero where coincidently the signal is strong, which then boosts the SNR at the low multipoles, and the total~SNR. \\

In light of these results, the optimal frequency for ISW detection appears to be around 353~GHz/850~$\mu m$ with a maximum SNR reaching 7$\sigma$. However in pratice, the CIB extraction at this frequency might prove challenging since the CMB becomes dominant here, and increasingly so as we go down in frequency. Therefore the possible residuals in our extracted CIB map have to be accounted for, and other sources of noise as well, which is the purpose of the next subsection.


\subsection{More realistic SNR}
\label{sec:real_noise}

We now carry a more realistic study by including several possible sources of contamination~: first the signal is completely dominated on a large part of the sky by emissions from our own galaxy. The contamination from this foreground in the galactic plane is several orders of magnitude above the CIB level and prevent us from extracting the CIB, therefore reducing the ``usable" fraction of the sky by at least $\sim25\%$. Furthermore, the rest of the sky is also quite polluted --~from a CIB point-of-view~-- by these foregrounds full of galactic dust. These will have to be removed from our maps although some residuals might remain in the final CIB map used for the cross-correlation. There may even be a significant CMB residual in this map, due to an imperfect separation of components. Consequently, we need to assess the impact of these contaminants in our study. \\

To account for these effects on the detectability of the CIB-CMB cross-correlation, we use in the present section a more realistic formulation of the signal-to-noise ratio, by adding new elements in the noise term. It therefore becomes at a given frequency $\nu$~:
\begin{multline}
\label{snrreal}
\left[\frac{S}{N}\right]^2 \! \! \! \! (\nu) = f_{\rm{sky}} \sum_{\ell=2}^{\ell_{\rm{max}}} (2\ell+1) \times \\ \frac{[C_{\ell}^{\rm{cr}}(\nu)]^2}{[C_{\ell}^{\rm{cr}}(\nu) \! + \! N_{\ell}^{\rm{cr}}(\nu)]^2+[C_{\ell}^{\rm{CIB}}(\nu) \! + \! N_{\ell}^{\rm{CIB}}(\nu)] [C_{\ell}^{\rm{CMB}} \! + \! N_{\ell}^{\rm{CMB}}]}
\end{multline}
where $f_{\rm{sky}}$ is the fraction of the sky common to the CMB and the CIB maps, and $N_{\ell}^{\rm{cr}}$, $N_{\ell}^{\rm{CIB}}$ and $N_{\ell}^{\rm{CMB}}$ are the noise contributions respectively in the cross, CIB and CMB signal. Since the CMB is expected to be only variance-limited at the multipoles of interest, we take here $N_{\ell}^{\rm{CMB}}=0$. However we still have to take into account the CIB contamination. \\

To do so, we first break the CIB noise power spectrum into several independant parts~:
\begin{equation*}
N_{\ell}^{\rm{CIB}}(\nu) = R_{\ell}^{\rm{CMB}}(\nu) + R_{\ell}^{\rm{fore.}}(\nu) + N_{\ell}^{\rm{instr.}}(\nu) + N_{\ell}^{\rm{correl.}}(\nu)
\end{equation*}
where these four different terms represent, from left to right, the power spectra of the CMB residual, the galactic foreground residuals, the instrumental noise and finally the noise due to correlation between residuals and the CIB (which appears when autocorrelating the final CIB map). \\ 

We quantify the CMB residual in the CIB map as a fraction $f_{\rm{CMB}}$ of the total CMB map, which affects both the cross-correlation and CIB noise. This consequently defines the noise in the cross signal~:
\begin{equation*}
N_{\ell}^{\rm{cr}}(\nu) = f_{\rm{CMB}}(\nu) \times C_{\ell}^{\rm{CMB}}
\end{equation*}
and the following two contributions~:
\begin{align*}
R_{\ell}^{\rm{CMB}}(\nu) &= f_{\rm{CMB}}^2(\nu) \times C_{\ell}^{\rm{CMB}} \\
N_{\ell}^{\rm{correl.}}(\nu) &= 2 f_{\rm{CMB}}(\nu) \times C_{\ell}^{\rm{cr}}(\nu) 
\end{align*}
We then define the spectrum of the foreground residuals as the following power law~:
\begin{equation*}
R_{\ell}^{\rm{fore.}}(\nu) = \mathcal{A}_{\rm{fore.}}(\nu) \times C_{\ell=10}^{\rm{CIB}}(\nu) \left(\frac{\ell}{10}\right)^{\alpha} 
\end{equation*}
so that their amplitudes are defined relatively to the real CIB signal through a chosen constant $\mathcal{A}_{\rm{fore.}}$, which defines the quantity~:
\begin{equation*}
\mathcal{A}_{\rm{fore.}}(\nu) = R_{\ell=10}^{\rm{fore.}}(\nu)/C_{\ell=10}^{\rm{CIB}}(\nu) \ ,
\end{equation*}
i.e. the ratio between the foreground residuals and the CIB spectrum at the multipole $\ell=10$, approximatively where the cross-signal is at its maximum. The slope of the spectrum $\alpha$ is fixed here for all frequencies ; previous analysis of infrared maps \citep{dust_spec,MAMD_dust} found it to be $\simeq-3$ for foregrounds at high galactic latitudes. Finally, the instrumental noise power spectra $N_{\ell}^{\rm{instr.}}$ at each frequency are taken from the first ten months of \textit{Planck} data in \citet{HFI_Proc}, and extrapolated to the thirty months, i.e. the end of the fourth \textit{Planck} full-sky survey. \\

In this section we focus on four of the five previously described \textit{Planck} HFI freqencies, from 217 to 857~GHz~: we discard the fifth 143~GHz as the CMB completely dominates the CIB signal there. We also put aside the IRAS frequency here because of its weaker significance, and the SPIRE frequencies since the instrument is not scheduled to ever cover very large regions of the sky (i.e. $f_{\rm{sky}} \ll 1$), dramatically decreasing the realistic SNR (see Eq.~(\ref{snrreal})). \\

At this point, we get three free parameters at each of the four frequencies in our SNR analysis : $f_{\rm{sky}}$, $f_{\rm{CMB}}$ and $\mathcal{A}_{\rm{fore.}}$. The next step would be to explore this 3D parameter space at each frequency and compute the SNR at each point. Considering the very large number of possible combinations of parameters, it would not be practical to display the complete results of this exploration here. Therefore we first choose to fix $f_{\rm{sky}}$ to two values of interest~: 
\begin{itemize}
\item $f_{\rm{sky}}=0.75$, which corresponds to an optimistic case where the only part of the sky that we discard is the galactic plane ; unfortunately there are other highly contaminated regions where the component separation techniques might not be able to extract the CIB.
\item $f_{\rm{sky}}=0.15$, which is a low estimate of the area of the sky where the current data allow for an efficient CIB extraction. The methods currently employed are based on the use of HI maps as a tracer of the galactic dust, though it only remains valid for an HI column density lower than a specific threshold \citep[see][for details on these methods]{CIB_Planck}. 
\end{itemize}
Concerning our other two parameters we limit ourselves to reasonable values, with $f_{\rm{CMB}} \in [0,0.1]$ and $\mathcal{A}_{\rm{fore.}} \in [0,10]$. \\

   \begin{figure}
   \centering
   \includegraphics[width=7cm]{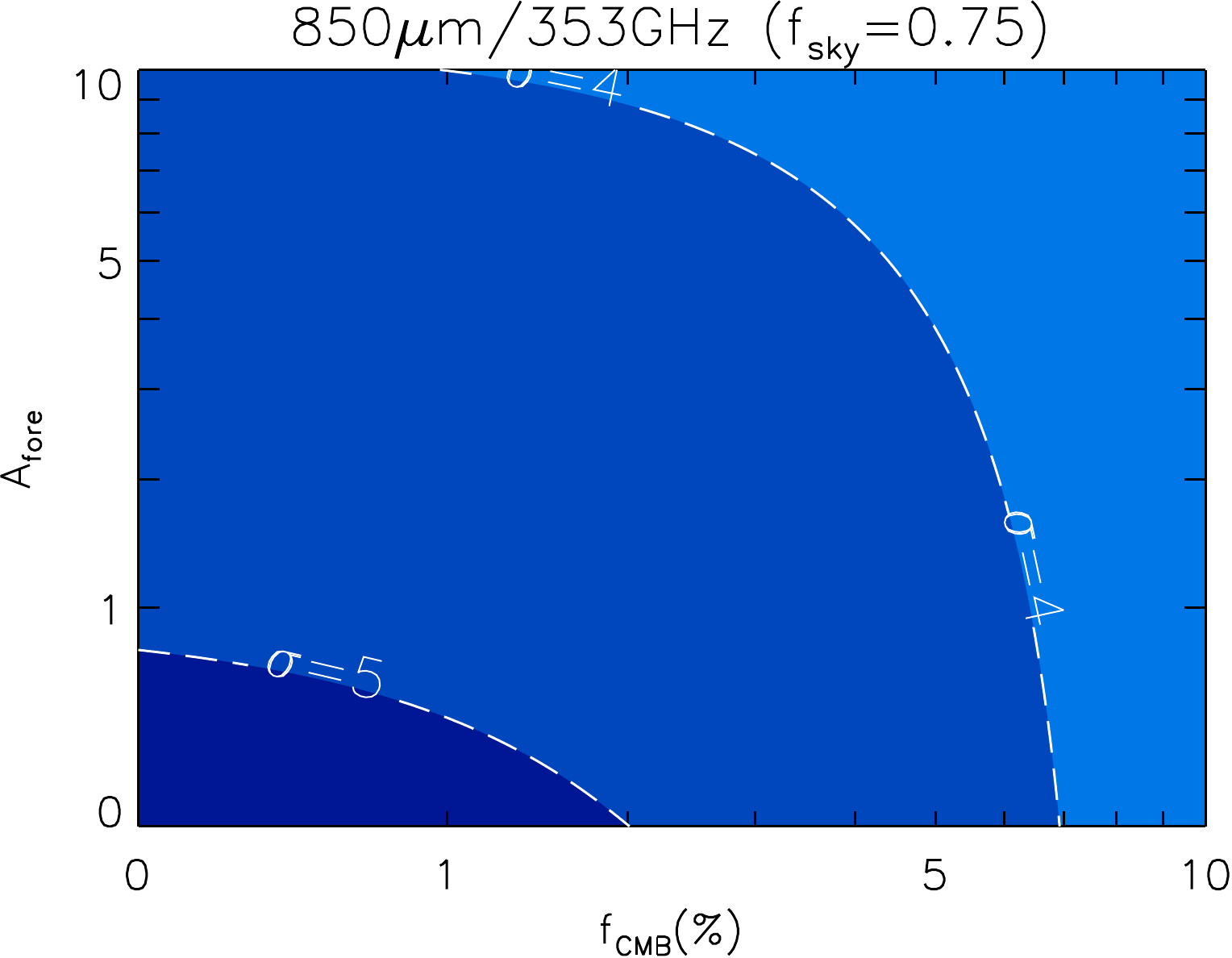} \\ \ \\
   \includegraphics[width=7cm]{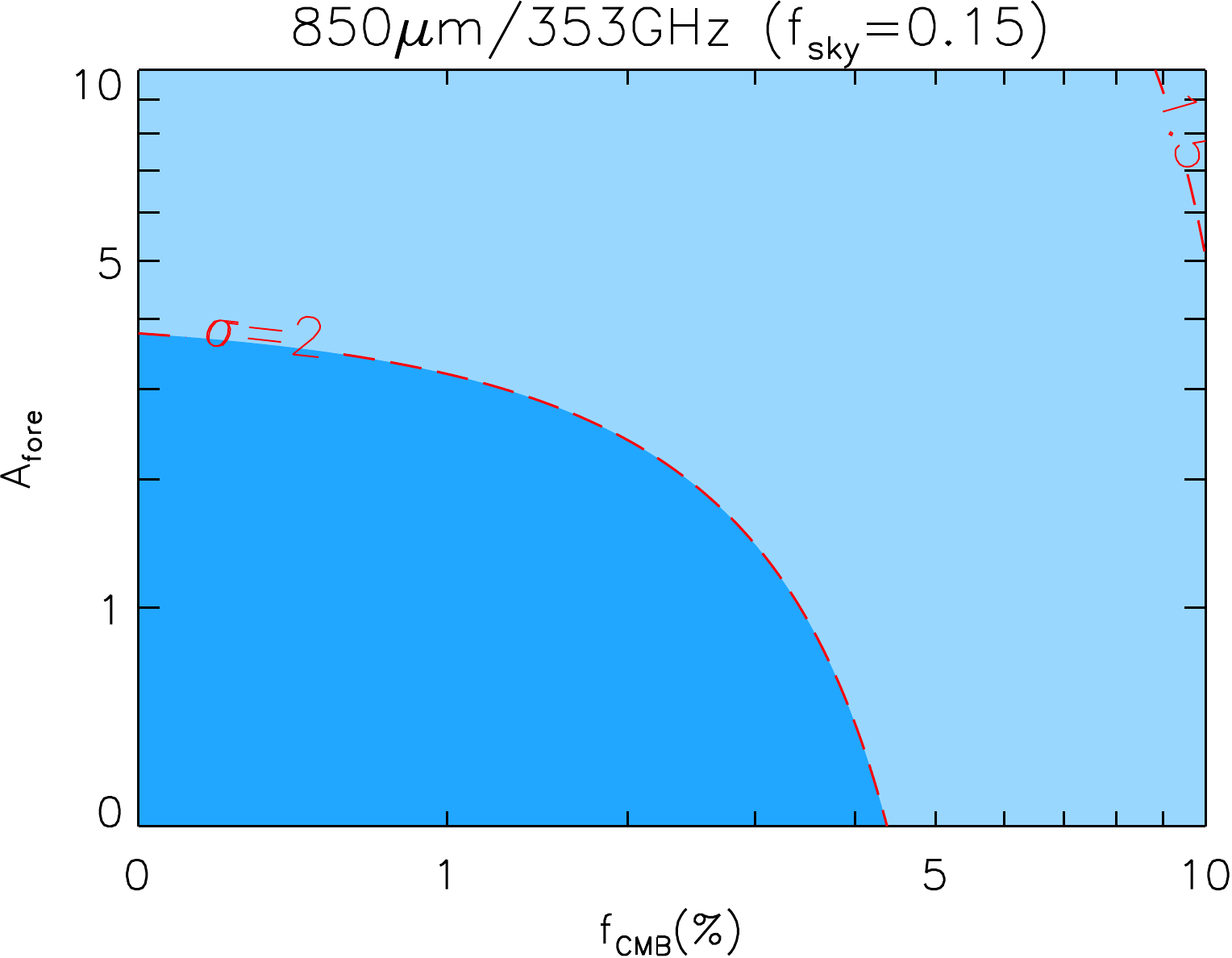}
      \caption{Total signal-to-noise ratio of the CIB-CMB cross-correlation at 353 GHz, as a function of the CMB residuals (in percentage of the total CMB signal) and the foregrounds residuals (through the parameter $\mathcal{A}_{\rm{fore.}}$). \textit{Upper panel :} $f_{\rm{sky}}=0.75$, the results go from less than 4 to more than 5, from the brightest colored area to the darkest. \textit{Lower panel :} $f_{\rm{sky}}=0.15$, the SNR goes from slightly less than 1.5 to more than 2, again from the brightest to the darkest area.}
         \label{snr_graph1}
   \end{figure}

We then focus on the frequency that gave the best SNR results in the ideal case, namely 850~$\mu m$/353~GHz, and study the effect of the noise on the cross-correlation detectability. The results are presented in Fig.~\ref{snr_graph1} which shows the contour levels of the SNR in the $(f_{\rm{CMB}},\mathcal{A}_{\rm{fore.}})$ parameter space. The influence of the CMB is clearly visible at this frequency, quickly reducing the SNR as its residual level increases. This effect is even more pronounced at 1380~$\mu m$/217~GHz, where the SNR is typically twice as low as in the ideal case (see Table~\ref{SNRtab}), due to the fact that we get closer to the maximum of the SED of the CMB. It makes this frequency far less significant for the ISW detection than in the ideal case. The presence of instrumental noise --~whose effect cannot be appreciated with Fig.~\ref{snr_graph1} alone~-- becomes significant at the two lowest frequencies (217 and 353~GHz), again reducing their value in the cross-correlation. As expected the galactic foreground residuals also decrease the SNR, though their influence is roughly the same at all frequencies as they are defined relatively to the CIB spectrum in our analysis. Lastly, the biggest influence comes from the fraction of the sky through the $f_{\rm{sky}}$ parameter, as the total SNR scales as $\sqrt{f_{\rm{sky}}}$. This makes it a crucial requirement for future applications to have the largest possible coverage to minimize this effect.\\

\begin{figure}
   \centering
   \includegraphics[width=7cm]{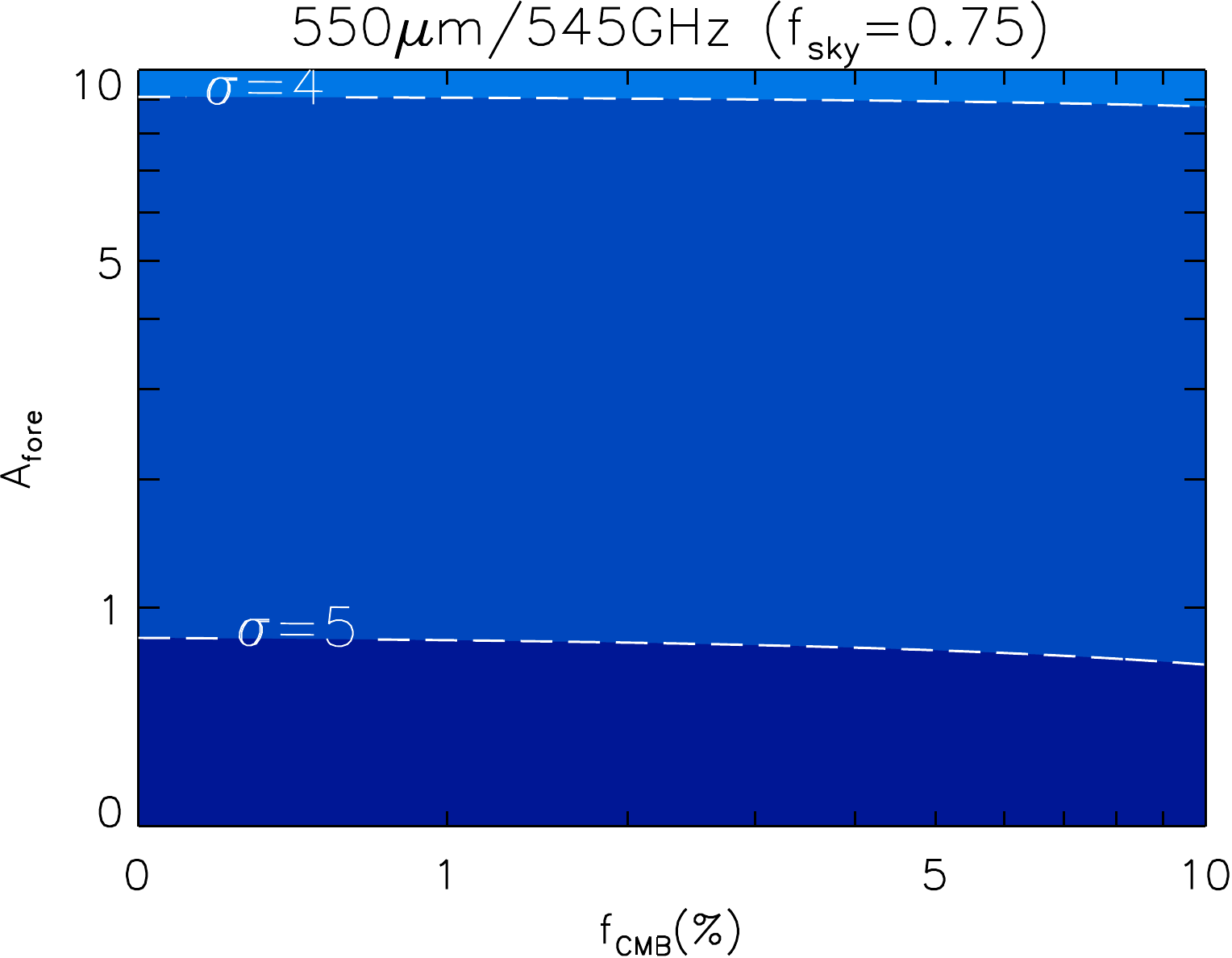} \\ \ \\
   \includegraphics[width=7cm]{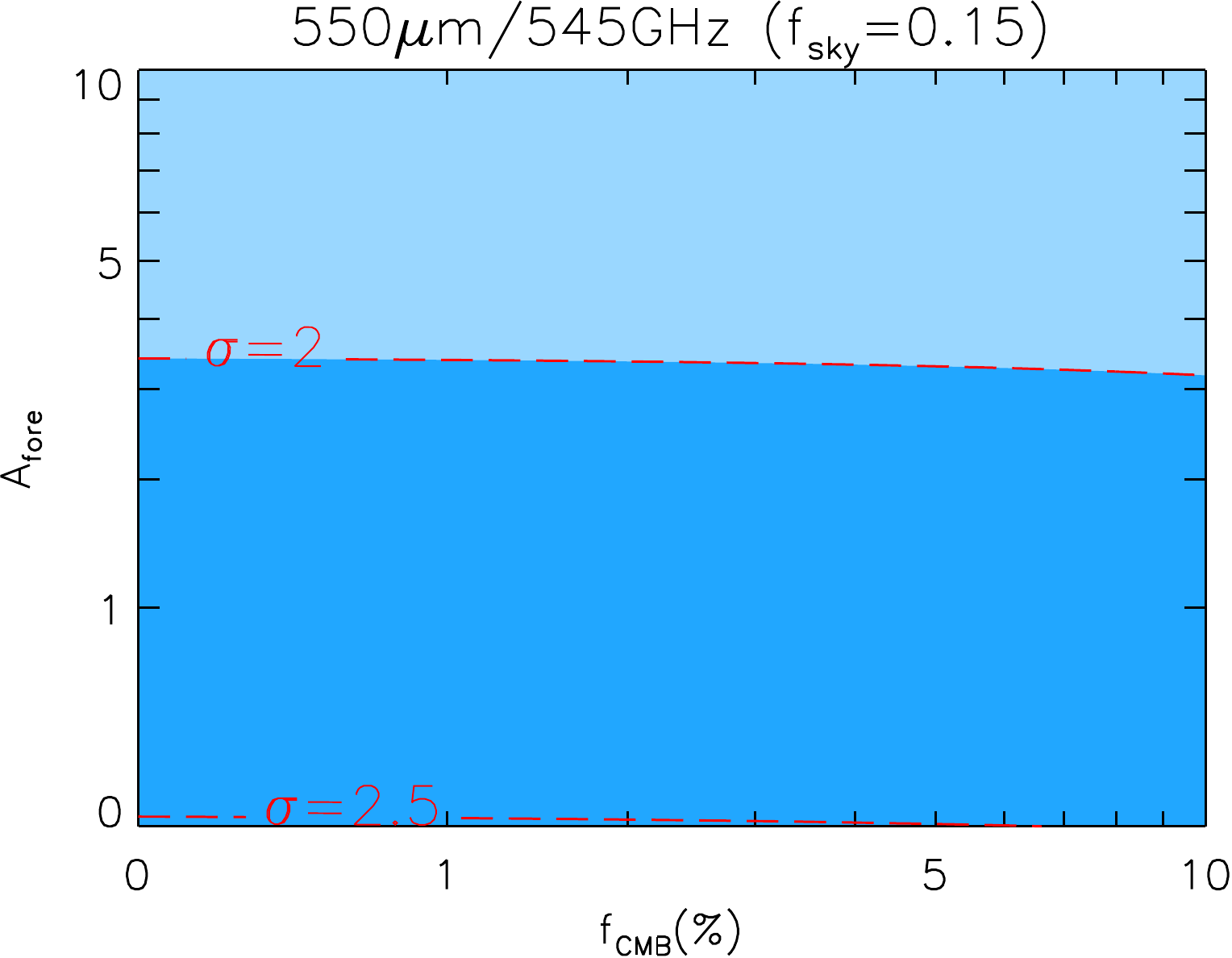}
      \caption{Total signal-to-noise ratio of the CIB-CMB cross-correlation at 545 GHz as a function of the CMB residuals and the foregrounds residuals. \textit{Upper panel :} $f_{\rm{sky}}=0.75$, the results go from slightly less than 4 to more than 5, from the brightest to the darkest area. \textit{Lower panel :} $f_{\rm{sky}}=0.15$, the SNR goes from less than 2 to slightly more than 2.5.}
         \label{snr_graph2}
   \end{figure}

\newpage
Taking all these remarks into account and after some exploration of the parameter space, the optimal frequency that stands out in these more realistic scenarii is 545~GHz/550~$\mu m$. Indeed, it is weakly influenced by instrumental noise and CMB residuals and also has a higher ``original" SNR (in the ideal case) than the other remaining frequency 857~GHz/350~$\mu m$. Our analysis at 545 GHz is presented in Fig.~\ref{snr_graph2}.

\subsection{Joint SNR}
\label{sec:joint}

Until now we have only considered a detection at a single CIB frequency and its associated significance. In practice, we will have several cross spectra at different frequencies, e.g. in the case of \textit{Planck} where we will be able to extract the CIB at four different frequencies on a large fraction of the sky. This allows us to increase the total signal-to-noise ratio of the ISW detection by combining the constraints from all available frequencies, though this will be limited by the possible intrinsic correlations between the CIB maps at different frequencies. Indeed, such correlations imply some redundancy in the measured information, and therefore lessen the gain in the total significance of the combined detection. \\

We can expand the previous SNR formalism to express the theoretical joint significance of a set of $n$ cross-correlations (i.e. CIB at $n$ frequencies, each correlated to the same CMB)~:
\beq{sntot}
\left( \frac{S}{N} \right)^2_{\rm{Total}}=X^T\mathcal{M}^{-1}X
\eeq 
with $X$ ($X^T$) being the column (row) vector of all the cross-correlations :
\begin{displaymath} 
X^T = \left( X^T(\nu_1) \ \cdots \ X^T(\nu_n) \right)  
\end{displaymath}
where $X^T(\nu_i)$ contains the cross-spectrum at the frequency $\nu_i$, from $\ell=2$ to 100 :
\begin{displaymath} 
X^T(\nu_i) = \left( C_{\ell=2}^{\rm{cr}}(\nu_i) \ \cdots \  C_{\ell=100}^{\rm{cr}}(\nu_i) \right) 
\end{displaymath}
The block matrix $\mathcal{M}$ is the covariance matrix, containing $n \times n$ blocks. Each one of them represents the covariance of two cross-spectra at different CIB frequencies, depending on the position of the block. At the $i$th line and $j$th column, the block $\mathcal{M}^{ij}$ is written as~:
\begin{center}
$\mathcal{M}^{ij}=\begin{pmatrix}
\mathcal{M}^{ij}_{\ell=2} & \  &  0 \\
\  & \ddots & \  \\
0 & \  & \mathcal{M}^{ij}_{\ell=100}
\end{pmatrix}$
\end{center}
The diagonality of $\mathcal{M}^{ij}$ comes from the assumption that the different multipoles are uncorrelated. In the noiseless case discussed in Section~\ref{sec:perf_sn}, the elements of each block can be expressed as follows~: 
\begin{align*}
\mathcal{M}_\ell^{ij} &= \rm{Covar}(C_{\ell}^{\rm{cr}}(\nu_i),C_{\ell}^{\rm{cr}}(\nu_j)) \\
&= \frac{C_{\ell}^{\rm{cr}}(\nu_i) C_{\ell}^{\rm{cr}}(\nu_j) + C_{\ell}^{\rm{CMB}} C_{\ell}^{\rm{crCIB}}(\nu_i,\nu_j)}{2\ell+1}
\end{align*}
We can see here the dependence on the aforementionned possible correlation between the CIB at frequency $\nu_i$ and the CIB at frequency $\nu_j$, through the cross-spectrum $C_{\ell}^{\rm{crCIB}}(\nu_i,\nu_j)$. To perform a more advanced analysis, we can easily modify this expression to account for the possible sources of noise discussed in the previous section. \\

Once again, the large number of possible combinations of noise parameters makes it unpractical to present a complete study of the joint correlation. Instead we focus on a few particular cases, motivated by what we found in Section~\ref{sec:real_noise}. A summary of our results on single and joint correlation is presented in Table~\ref{SNRtab}~: we first go back to the ideal case to quantify the impact of the joint detection. We found a relatively small gain, as it increases the total SNR by a mere $\simeq 0.15$ compared to the maximum significance of a single detection. This can be attributed to the high correlations between the CIB at its different observed frequencies, which limit the usefulness of the joint cross-correlation.\\

Considering now more realistic situations, with the presence of instrumental noise, we once again choose to fix some of the parameters mentionned in Section~\ref{sec:real_noise}, with $f_{\rm{sky}}=0.75$ and $f_{\rm{sky}}=0.15$. A reasonable confidence in compenent separation techniques allows us to hope for small enough residuals, so that we choose $f_{\rm{CMB}}=0.01$ and $\mathcal{A}_{\rm{fore.}}=0.01$. In these cases, the joint correlation has once again a limited interest (respectively a $\simeq 0.15$ and $\simeq 0.07$ gain for $f_{\rm{sky}}=0.75$ and $0.15$) due to the correlations in both the CIB signals but also in the astrophysical noise contributions --~CMB and dust~-- between frequencies.

\begin{table}
\small
\begin{tabular}{|c||c|c|c|c|}
\hline
\textbf{Frequency (}GHz\textbf{)} & \textbf{857} & \textbf{545} & \textbf{353} & \textbf{217} \\
\hline
\textbf{Wavelength ($\mu m$)} & \textbf{350} & \textbf{550} & \textbf{850} & \textbf{1380} \\
\hline
\hline
\textbf{Perfect Single SNR} & 6.26 & 6.83 & 6.98 & 6.95 \\
\hline
\textbf{Joint SNR} & \multicolumn{4}{c|}{7.12} \\
\hline
\hline
\textbf{Realistic Single SNR 1} & \ & \ & \ & \ \\
($f_{\rm{sky}}=0.75$, $f_{\rm{CMB}}=0.01$, & 5.36 & 5.73 & 5.39 & 3.56 \\
$\mathcal{A}_{\rm{fore.}}=0.01$) & \ & \ & \ & \ \\
\hline
\textbf{Joint SNR} & \multicolumn{4}{c|}{5.88} \\
\hline
\hline
\textbf{Realistic Single SNR 2} & \ & \ & \ & \ \\
($f_{\rm{sky}}=0.15$, $f_{\rm{CMB}}=0.01$, & 2.40 & 2.56 & 2.41 & 1.59 \\
$\mathcal{A}_{\rm{fore.}}=0.01$) & \ & \ & \ & \ \\
\hline
\textbf{Joint SNR} & \multicolumn{4}{c|}{2.63} \\
\hline
\end{tabular}
\caption{Total signal-to-noise ratio of the CIB-CMB cross-correlation for four of the CIB frequencies of \textit{Planck}-HFI. The results are given for each frequency and for the joint cross-correlation, first for the ideal case discussed in Section~\ref{sec:perf_sn} and then for two more realistic cases.}
\label{SNRtab} 
\end{table}


\section{Conclusions}
\label{sec:conclusions}

The topic of this paper is an investigation of the cross-correlation between the cosmological infrared and microwave backgrounds, and a study of its detectability under various observational situations. A non-zero correlation is expected to exist between the two backgrounds and their anisotropies through the ISW effect, caused by the time-evolving gravitational potentials that underlie the large-scale structures which are the sources of the CIB and of its anisotropies. Describing the CIB anisotropies as linearly biased tracers of the matter field fluctuations, we calculated the theoretical angular power spectrum of the CMB-CIB cross-correlation at several frequencies and for different instruments, taking into account their actual bandpasses. As is well known for CMB/galaxies cross-correlations, the signal peaks at low multipoles and quickly vanishes at higher $\ell$. The linear bias introduced by our formalism was then obtained by confronting our predicted linear CIB power spectra with the data coming from the \textit{Planck} mission. These observed CIB anisotropies were fitted in \citet{CIB_Planck} by an HOD model, to which we compared our own spectra at the low multipoles in order to get the desired bias at each different frequency. \\

Using an advanced SNR analysis which included the main sources of noise both instrumental and astrophysical, and all their possible correlations, we pointed out the most promising frequency in the ideal case of noiseless full-sky maps (850~$\mu m$/353~GHz) with an expected significance as high as $\simeq 7\sigma$ for the cross-correlation signal. The same frequency turned out to be less optimal with more realistic assumptions about sky coverage and possible sources of noise (here CMB, dust residuals and instrumental noise). In this case, higher frequencies such as \textit{Planck}-HFI's 545 and 857 GHz are favored, with an expected significance ranging from 2.4 to 5.7 depending on the frequency, the levels of noise and the fraction of the sky available for analysis. We also found that a joint cross-correlation using all available frequencies is of minor interest, due to the high correlations between CIB anisotropies at the different frequencies. Nevertheless, our best results for $f_{\rm{sky}}=0.75$ are higher than the significances of all current CMB-galaxies cross-correlation, with $\sigma>5$, although a less optimistic estimate for the sky coverage quickly reduces our signal-to-noise ratios. This stresses once again the requirement of good component separation techniques and foreground removals for future applications, in order to have the largest fraction of common clean sky $f_{\rm{sky}}$ possible. \\

The results of this work will be valuable in the forthcoming years of analysis and exploitation of the \textit{Planck} data. The formalism we developped provides us with an accurate forecast of the expected results of the CIB-CMB cross-correlation and allows us to constrain the requirements for a significant ISW detection. Regarding the use of the CIB itself, it presents some advantages over classical ISW studies : the underlying structures observed through the CIB span a large integrated range of redshifts and cover the whole sky whereas the usual galaxy surveys used often have a limited depth and width in redshift or a small sky coverage --~some of the main limiting factors in the ISW detection. \\

In current studies, the CIB is always considered in its integrated form at a given frequency, meaning that in this observed CIB, contributions from many redshift ranges are mixed together. An interesting further step would be to use the multiple observed frequencies to reconstruct the contributions from different redshift bands, in order to obtain several decorrelated CIB maps corresponding to these redshift slices. The resulting independent CIB maps could then be individually correlated with the CMB. Combining the independent detections could increase even more the total SNR of the ISW detection, as it allows to get rid of the correlation terms between CIB maps. Furthermore, each of these maps will then help tracing the Dark Energy at a different time. Our preliminary calculations from predicted power spectra indicate encouraging enhancements in the signal-to-noise ratio, although the details of the CIB decorrelation need further investigation and optimization, and will be presented in a future work. \\

Finally, let us note that as a background the CIB is likely to be lensed by large-scale structures in the local Universe : a dedicated study of the effects of lensing in a future work will be able to determine if the lensing could lead to a possible gain in the signal-to-noise ratio of the ISW effect, or should be considered as a possible source of bias in the DE detection.

\section*{Acknowledgements}
We would like to thank Nabila Aghanim and Fabien Lacasa for valuable comments and fruitful discussions. We also thank Olivier Dor\'e for his encouraging comments. M.~L. and S.~I. acknowledge financial support by the Doctoral Programme ``AAP 2010 contrats doctoraux Paris-Sud 11".

\bibliographystyle{mn2e}
\bibliography{article}

\end{document}